\documentclass[english,american]{article}
\usepackage[T1]{fontenc}
\usepackage[latin9]{inputenc}
\usepackage[a4paper]{geometry}
\usepackage{babel}

\usepackage{amsmath}
\usepackage{amssymb}
\usepackage{stackrel}
\usepackage{graphicx}
\usepackage{setspace}
\usepackage[square,numbers]{natbib}
\usepackage{multicol,caption}
\usepackage{authblk}
\usepackage{color}
\usepackage{mathpazo}
\newenvironment{Figure}
  {\par\medskip\noindent\minipage{\linewidth}}
  {\endminipage\par\medskip}

\def\be{\begin{equation}}
\def\ee{\end{equation}}

\def\ds[#1]{\textcolor{blue}{#1}}

\begin{document}


\title{Voltage preservation in Intermediate Band Solar Cells}

\author[1,2]{Daniel Suchet\footnote{Email daniel.suchet@polytechnique.org}}
\author[1,2]{Amaury Delamarre}
\author[2,3]{Nicolas Cavassilas}
\author[1,2]{Zacharie Jehl}
\author[1,2]{Yoshitaka Okada}
\author[1,2]{Masakazu Sugiyama}
\author[1,4]{Jean-Francois Guillemoles}
\affil[1]{Research Center for Advanced Science and Technology, The University of Tokyo, 4-6-1 Komaba, Meguro-ku, Tokyo 153-8904, Japan}
\affil[2]{LIA NextPV, Research Center for Advanced Science and Technology, The University of Tokyo, 4-6-1 Komaba, Meguro-ku, Tokyo 153-8904, Japan}
\affil[3]{Aix Marseille Universit\'e, CNRS, Universit\'e de Toulon, IM2NP UMR 7334, 13397, Marseille, France}
\affil[4]{Institut Photovoltaique d'Ile de France (IPVF), UMR 9006 , 30 route d\'epartementale 128, 91120, Palaiseau, France}
\maketitle

\begin{abstract}
Intermediate Band Solar Cell is an advanced concept for solar energy conversion in which two low-energy photons can promote an electron to the conduction band through a so-called \emph{intermediate band}. To limit recombination and preserve the photo-generated voltage, generation to- and from the intermediate band should be matched. However, all practical realizations experienced a significant voltage degradation as compared to a single junction without intermediate band. In this work, we develop a novel analytical optimization method based on Lagrange multipliers. We demonstrate that an Intermediate Band Solar Cell under solar spectrum cannot meet voltage preservation and current matching at the same time. By contrast, we show that the implementation of an energy shift (\emph{electronic ratchet}) in any of the bands allows those two criteria to be filled simultaneously. Additional insights are provided by the numerical study of the short circuit current and fill factor of the systems at stake, which show that a system with ratchet benefits from the same current increase as a standard Intermediate Band Solar Cell (same short-circuit current), while maintaining I-V properties of a single junction (same open-voltage circuit, same fill factor).
\end{abstract}

\begin{multicols}{2}
\section{Introduction}
Intermediate band solar cells were introduced as a new concept of photovoltaic converter to overcome the celebrated Shockley Queisser limit \cite{shockley_detailed_1961}. IBSC aims at collecting some photons with energy below the energy bandgap, that would not be absorbed in a standard single- junction solar cell. To do so, a so-called intermediate band is introduced within the bandgap and serves as built-in up-converter, allowing two photons transitions from the valence band to the intermediate band (IV transition) and from the intermediate band to the conduction band (IC transition). This additionnal absorption increases the current produced by the device, and allows theoretically for an efficiency enhancement from 31\% to 47\% under one sun illumination \cite{luque_increasing_1997}. 

IBSC have received large attention over the last decades, and several refinement have been brought to the seminal concept \cite{luque_thermodynamic_2001, cuadra_influence_2004, levy_solar_2008, levy_intraband_2008}. However, despite experimental efforts (see \cite{okada_intermediate_2015} and references therein), no fully working proof of concept has been reported yet. One of the main blocking point is certainly the voltage degradation induced by recombination through the intermediate band \cite{linares_voltage_2012}. Furthermore, it was recently suggested that the combined effect of small non-idealities, such as narrow absorption and non radiative recombination on the IC transition could prevent IBSC from reaching SQ limit \cite{delamarre_electronic_2017}. An elegant way to circumvent this issue is to introduce an \emph{electronic ratchet}, i.e. an energy shift between the IV and IC transitions \cite{yoshida_photon_2012}. Possible realizations could rely on succession of quantum wells organized with tailored width \cite{delamarre_electronic_2017} or as a quantum cascade \cite{curtin_quantum_2016}, on 2D materials \cite{chen_design_2017} or on compounds with transition metal \cite{olsson_ferromagnetic_2009}.  It has been shown numerically that such \emph{electronic ratchet} not only increases the conversion efficiency of IBSC, but also strongly enhances their resilience against the aforementioned non-idealities. Despite the increasing interest for the ratchet feature \cite{sahoo_use_2018}, the physics at stake behind the ratchet system is not trivial \cite{pusch_limiting_2016, bezerra_lifetime_2017}, and received little attention so far. 

While conversion efficiency is a well suited indicator for practical purposes of energy conversion, it is a too aggregated figure of merit to provide a clear picture of the conversion processes. This is especially true for complex systems such as IBSC, where many processes take place at the same time \cite{yoshida_photon_2012}. On the other hand, electrical properties of the device, such as open circuit voltage, short circuit current and fill factor, offer a deeper perspective. Most notably, being directly related to the carriers free energy, open circuit voltage has been proven to bring insight on the system thermodynamics \cite{wurfel_chemical_1982}, and can be used to estimate the entropy production occuring during the photon-to-carrier conversion  \cite{markvart_thermodynamics_2007}. Studying open-circuit voltage also sheds light on the voltage preservation issue that prevents standard IBSC from exceeding the Shockley Queisser limit.

In this work, we study the open circuit voltage of a IBSC with electronic ratchet (RBSC). Our approach relies on Lagrange multipliers, which allows to estimate the optimal configuration of a system under constraints. Using this powerful technique, we show analytically that an optimal RBSC displays the same $V_{{\rm OC}}$ as a single junction with the same energy gap. This property can be used to estimate the optimal band configuration. By contrast, we show that IBSC can not preserve voltage and ensure a good current matching at the same time. Finally, complementary figures of merit are numerically studied, suggesting physical interpretation of the ratchet influence.

\begin{Figure}
 \centering
 \includegraphics[width=\linewidth]{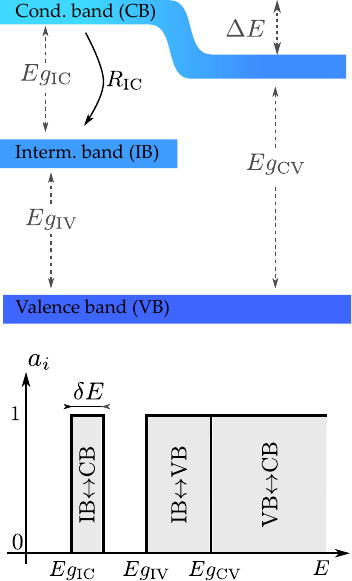}
 \captionof{figure}{Band configuration and absorptivity in a RBSC. In the non-overlapping absorption model, photons with between $Eg_{IC}$ and $Eg_{IC}+\delta E$ are absorbed and emitted in the intermediate to conduction (IC) transition ; those with energy between $Eg_{IV}$ and $Eg_{CV}$ in the IV transition and those with energy above $Eg_{CV}$ in the CV transition. Non radiative losses are considered on the IC transitions, in addition to radiative recombinations. We account for this effect through a radiative effiency factor increasing $R_{IC}$ (pictured with a curved arrow).}
\label{Fig:system}
\end{Figure}
\section{Model and notations}
We will use the detailed balance model introduced and detailed by
Yoshida \emph{et al.} \cite{yoshida_photon_2012}, with notations presented in Fig \ref{Fig:system}. The
carrier generation $G_{i}$ and radiative recombination $R_{i}$ rates
for each transition i are given by a generalized Planck's
law \cite{wurfel_chemical_1982}
\begin{align}
G_{i} &=\frac{f}{4\pi^{3}\hbar^{3}c^{2}}\int_{0}^{\infty}dE\,\frac{a_{i}(E)\, E^{2}}{\exp \left(\frac{E}{kT_S}\right) -1} \\
R_{i} &=\frac{1}{r_{{\rm rad,\,i}}}\frac{1}{4\pi^{2}\hbar^{3}c^{2}}\int_{0}^{\infty}dE\,\frac{\epsilon_{i}(E)\, E^{2}}{\exp \left(\frac{E-\Delta\mu_{i}}{kT_c}\right)-1}
\end{align}
where $i=VC,\,IV,\,IC$ labels the valence to conduction, valence to intermediate and intermediate to conduction band transitions respectively. $f$ is a geometrical concentration factor, equal to $6.79\times10^{-5}$ for an un-concentrated illumination.
The temperature $T$ is taken to $T_{S}=6000$ K for generation rates (dark current is neglected) and $T_{C}=300$ K for recombination rates, while the quasi Fermi
level splitting $\Delta\mu_{i}$ are all zero for the solar black-body radiation.
To ease notations, we will note $\Delta\tilde{\mu_{i}}=\Delta\mu_{i}/kT_{C}$.
The absorptivity, emissivity and radiative efficiency of each transition are denoted
$a_{i}(E)$, $\epsilon_{i}(E)$ and $r_{{\rm rad,\,i}}$ respectively. Owing to the Kirchhoff's law of radiation, emissivity is equal to absoptivity for each wavelength, and both generation and recombination rates can be inferred from $a_i$ only.  Following previous models, we will consider non-overlapping perfect absorptivity ($a_{i}(E)=1$) and radiative limit ($r_{{\rm rad,\,i}}=1$) for VC and IV transitions, while the IC transition is considered with a narrow span $\delta E$ and includes non-radiative recombination (see Fig. \ref{Fig:system}). This approach accounts for constraints raised by the use of nanostructures for IBSC, where the IB - CB transition is intra-band \cite{delamarre_electronic_2017, cavassilas_beneficial_2018}. 

The energy shift of the electronic ratchet is defined as 
\begin{equation}
Eg_{IV}+Eg_{IC}=Eg_{CV}+\Delta E.
\end{equation}
Note that the ratchet shift can be indifferently considered on the conduction band (as is the case here) or on the intermediate band \cite{pusch_limiting_2016}. In the following, we will refer to systems with finite $\Delta E$ as Ratchet Band Solar Cells (RBSC) and systems with $\Delta E=0$ as
IBSC - even though both systems rely on electronic transition through
an intermediate band. 

Electrically, these systems are equivalent to two diodes connected in series (IC and IV transitions), in parallel to a third diode (CV transition). As clear from this picture, Kirchhoff's voltage
and current laws impose constraints on the system, and can be written
as
\begin{align}
qV=\Delta\mu_{CV} & =\Delta\mu_{IV}+\Delta\mu_{IC}\label{eq:maille}\\
G_{IV}-R_{IV} & =G_{IC}-R_{IC}\label{eq:noeuds}
\end{align}
To simplify calculations, we will consider the Boltzmann approximation
of the recombination rates
\begin{equation}
R_{i}\simeq R_{i}^{0}\,\exp\left(\frac{\Delta\mu_{i}}{kT_{C}}\right)\label{eq:approxR}
\end{equation}
which is valid for all gaps both at open circuit voltage $V_{{\rm OC}}$
and maximum power point $V_{{\rm m}}$ under un-concentrated illumination. 

In the following, we will often consider the \emph{best-gap voltage} $V_{{\rm gm},\,i}$, defined as the bias for which an isolated junction with an energy gap $Eg_i$ would be optimal
\begin{equation}
\left. \partial_{Eg_{i}}\left[V\,\left(G_{i}-R_{i}(V)\right)\right]
\right|_{V=V_{{\rm gm},\,i}}=0
\end{equation}
The best-gap voltage is determined by the absorptivity, and thus depends only on the energy gap and absorption width. Within Boltzmann approximation (\ref{eq:approxR}), $V_{{\rm gm},\,i}$ can be expressed as:
\begin{equation}
\exp\left(\frac{qV_{{\rm gm},\,i}(Eg,\,\delta E)}{kT_{C}}\right)\simeq\frac{\left.\partial_{Eg_{i}}G_{i}\right|_{Eg_{i}=Eg}}{\left.\partial_{Eg_{i}}R_{i}^{0}\right|_{Eg_{i}=Eg}}.
\end{equation}

The best-gap voltage should not be confused with the maximum power point  $V_{\rm m}$, defined for a given absorptivity as
\begin{equation}
\left. \partial_{V}\left[V\,\left(G-R(V)\right)\right]
\right|_{V=V_{\rm m}}=0
\end{equation}

Considered at any value of the energy gap, the best-gap voltage is
\emph{a priori} different from the maximum power point, and both quantities coincide only for the optimal configuration. However, for a single junction, the best-voltage gap has been proven to be numerically close to the maximum power point down to few percents \cite{hirst_fundamental_2011}, and we will therefore consider
\begin{equation}
V_{{\rm gm},\,CV}\simeq V_{{\rm m}}^{(1)}\label{eq:Vm(1)}
\end{equation}
where $V_{{\rm m}}^{(1)}$ is the maximal power point of a single junction with gap $Eg_{CV}$.

\section{Voltage preservation in RBSC}

Following the model presented in the previous section, the efficiency
of RBSC system can be estimated from six parameters $\{Eg_{i},\,\Delta\mu_{i}\}$,
submitted to two constraints eq. (\ref{eq:maille}) and (\ref{eq:noeuds}).
In this section, we will study the open circuit voltage of an optimal
RBSC, ie for a configuration such that the power output power of the
cell
\begin{equation}
P\left(E_{gi},\,\Delta\mu_{i}\right)=\underset{i}{\sum}\Delta\mu_{i}\left(G_{i}-R_{i}^{0}e^{\Delta\tilde{\mu}_{i}}\right)
\end{equation}
is maximum. An appropriate framework to account for such a system
is provided by Lagrange multiplier \cite{goldstein_classical_2002}, where all parameters are treated as independent, and constraints are included explicitly in
the Lagrangian 
\begin{align}
L_{{\rm R}} =P+ & \lambda_{J}\left[\left(G_{IV}-R_{IV}\right)-\left(G_{IC}-R_{IC}\right)\right] \nonumber \\ 
& +\lambda_{V}\left[\Delta\mu_{CV}-\left(\Delta\mu_{IV}+\Delta\mu_{IC}\right)\right]\label{eq:LR}
\end{align}

Note that this approach can be extended to standard IBSC by adding a constraint on the energy gaps 
\begin{align}
L_{{\rm IBSC}} = L_{{\rm R}}+\lambda_{E}\left[Eg_{CV}-\left(Eg_{IV}+Eg_{IC}\right)\right]
\end{align}
which suggests that the benefit of the electronic ratchet comes from the relaxation of this additional constraint. 

The optimal configuration for the RBSC is reached when derivatives of the Lagrangian (\ref{eq:LR}) with respect to the system parameters $\{Eg_{i},\,\Delta\mu_{i}\}$ and Lagrange multipliers $\{\lambda_{J},\,\lambda_{V}\}$ are all equal to zero. The detailed calculations are presented in annex, and provide relations between parameters in optimal configuration.

It is notably possible to express the maximum power point of an optimal RBSC $V_{{\rm m}}=\Delta\mu_{CV}/q$ as a function of the three energy gaps only 
\begin{align}
\exp\left(\frac{qV_{{\rm m}}}{kT_{C}}\right)  = \exp\left(\frac{qV_{{\rm gm},\,CV}}{kT_{C}}\right)
\left(1+\frac{R_{IC}^{0}}{R_{IV}^{0}}
\frac
{e^{\frac{qV_{{\rm gm},\,IC}}{kT_{C}}}-1}{e^{\frac{qV_{{\rm gm},\,IV}}{kT_{C}}}}\right)^{-1}\label{eq:Vm-full}
\end{align}

The right hand side denominator can be shown to be close to unity. Indeed, to optimize current matching through the intermediate band and limit recombination, generation rates $G_{IC}$ and $G_{IV}$ should be equal (see section \ref{sec:opt}). However, a too narrow absorption width $\delta E$ can prevent $G_{IC}$ from reaching values close to $G_{IV}$. In any case, $G_{IV}$ forms an upper bound for the optimal value of $G_{IC}$. Due to current conservation eq.(\ref{eq:maille}), $R_{IC}$ is therefore also limited by $R_{IV}$, and is much small than this upper bound for narrow transitions. We will thus consider
\begin{equation}
R_{IC}^{0}\,e^{\frac{qV_{{\rm gm},\,IC}}{kT_{C}}} \leq R_{IV}^{0}\,e^{\frac{qV_{{\rm gm},\,IV}}{kT_{C}}} \label{eq:approx}
\end{equation}

With an estimation error smaller than $\log 2 \, kT_C / q $ (less than $2 \%$ of relative value), the previous expression can be simplified to
\begin{equation}
V_{{\rm m}} \simeq V_{{\rm gm},\,CV} \simeq V_{{\rm m}}^{(1)}\label{eq:mu1}
\end{equation}
where the last equality follows from eq. (\ref{eq:Vm(1)}). The maximum power point for a RBSC is close to that of single junction with the same large gap $Eg_{CV}$.

Furthermore, it can also be shown (see Appendix) that the open circuit voltage $V_{{\rm OC}}$ and maximum power point $V_{{\rm m}}$ for
a RBSC are related by the exact same relation as for a single junction:
\begin{equation}
\left(1+\frac{qV_{{\rm m}}}{kT}\right)\exp\left(\frac{qV_{{\rm m}}}{kT_{C}}\right)=\exp\left(\frac{qV_{{\rm OC}}}{kT_{C}}\right)\label{eq:Vm-Voc}
\end{equation}
As this relation is monotonous for positive biases, equation (\ref{eq:mu1})
implies the equality of open circuit voltage between the optimal RBSC
and the corresponding single junction with the same large gap:
\begin{equation}
V_{{\rm OC}}\simeq V_{{\rm OC}}^{(1)}. \label{eq:Voc-RBSC}
\end{equation}
This proves approximate voltage preservation in an optimal RBSC. This remarkable result also implies that, in an optimal RBSC, all three transitions reach their respective open-circuit condition $G_i = R_i$ simultaneously.

The open-circuit voltages of RBSC and IBSC are compared to that of a single junction sharing the same gap $Eg_{CV}$ on Fig. \ref{Fig:Voc}. For illustration purpose, we have taken an absorption width $\delta E = 150 \, {\rm meV}$ \cite{georgiev_short-wavelength_2003} and a radiative efficiency $r_{{\rm rad},IC}$ equal to $1$ or $10^{-3}$ \cite{yang_infrared_1995}. For small $Eg_{CV}$, an IBSC is limited to Shockley Queisser limit. No current flows through the intermediate transition, and the IBSC behaves like a single junction. At larger gaps, the intermediate transition contributes to the current, increasing the system efficiency. However, this transition also results in a loss larger than 10\% of the open circuit voltage as compared to the single junction. By contrast, the open-circuit voltage RBSC remains very close to $V_{{\rm OC}}^{(1)}$ (less than 1\%), regardless of the main gap or radiative efficiency considered.

It should also be noted that, due to the Lagrange multiplier approach used for its derivation, voltage preservation (\ref{eq:Voc-RBSC}) is only valid for optimal set of parameters $\{Eg_{i},\,\Delta\mu_{i}\}$. However, its numerical validity range appears to be significantly larger, and holds for a large span of gaps $Eg_{CV}$. Therefore, RBSC are expected to exhibit voltage preservation even in non-optimal situations.
\begin{Figure}
 \centering
 \includegraphics[width=\linewidth]{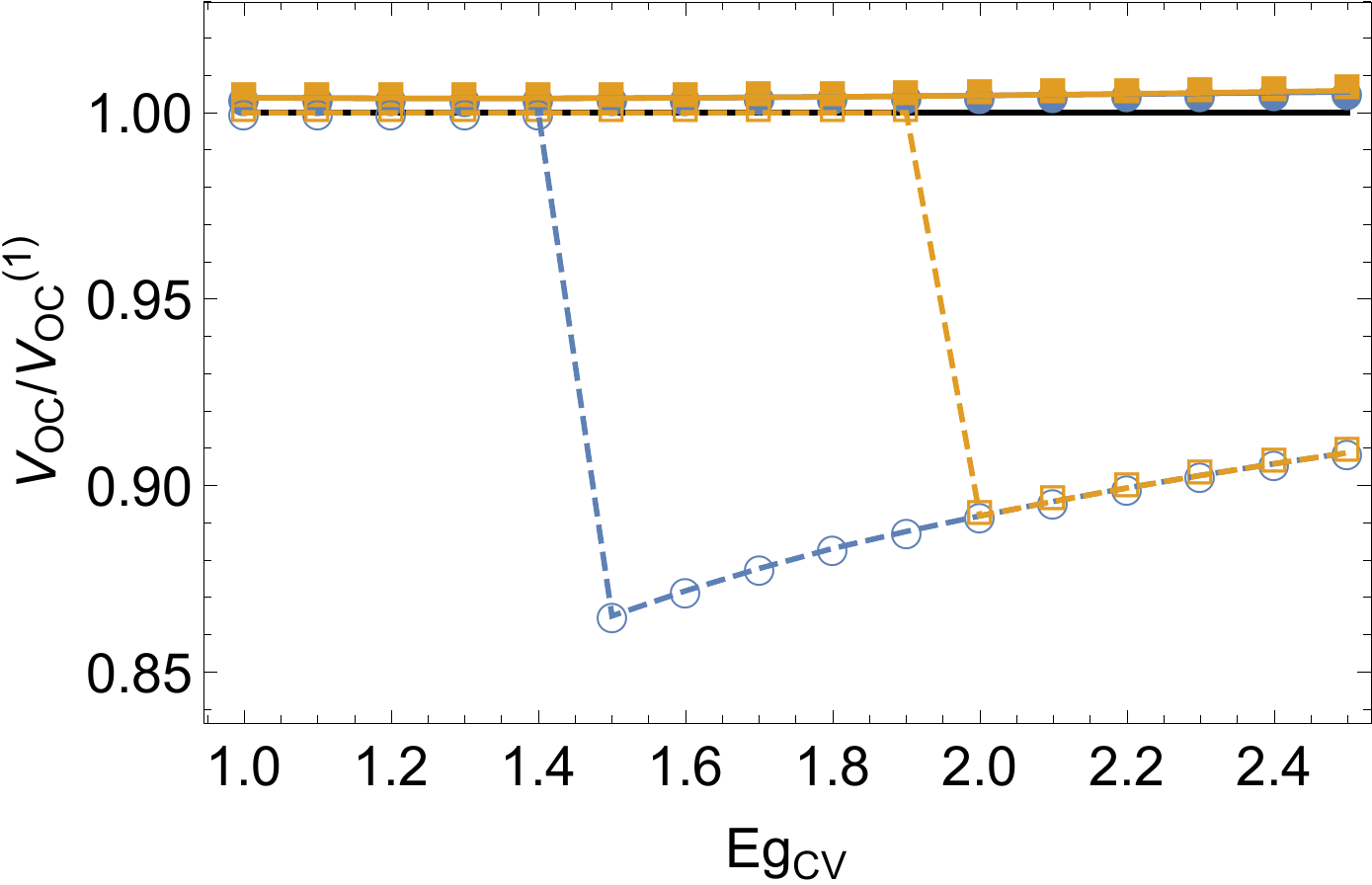}
 \captionof{figure}{Open circuit voltage of RBSC (plain line and filled symbols) and IBSC (dashed line and empty symbols), normalized to that of a single junction with the same main gap $Eg_{CV}$. Absorption width is taken to $\delta E = 150 \, {\rm meV}$ and $r_{{\rm rad},IC}$ to $1$ (blue circles) and $10^{-3}$ (yellow square). For each gap $Eg_{CV}$, the band configuration of both structures are determined by optimizing the conversion efficiency.}
\label{Fig:Voc}
\end{Figure}
\section{Optimization criteria for RBSC}\label{sec:opt}

In previous works, RBSC band configuration was optimized numerically by testing all possible combinations of parameters matching Kirchhoff laws eq.(\ref{eq:maille}) and (\ref{eq:noeuds}). This brute force method can be time consuming, but requires no physical understanding of the system. The Lagrange multiplier approach presented in the previous section provides additional constraints
(see eq.(\ref{eq:Vm-full}) and eq.(\ref{eq:muIC}-\ref{eq:muIV}) in annex), reducing the parameter
space. A sixth expression can be derived from the Lagrangian $L_{{\rm R}}$,
providing a closure relation on the electrical current $J_{{\rm m}}$
at the maximum power point:
\begin{align}
\frac{J_{{\rm m}}}{q} &=\frac{qV_{{\rm m}}}{kT_{C}}\exp\left(\frac{qV_{{\rm m}}}{kT_{C}}\right)\nonumber \\
&\times \left(R_{CV}^{0}+\frac{R_{IV}^{0}R_{IC}^{0}}{R_{IC}^{0}\,e^{\frac{qV_{{\rm gm},\,IC}}{kT_{C}}} + R_{IV}^{0}\,e^{\frac{qV_{{\rm gm},\,IV}}{kT_{C}}}}\right)
\end{align}

These relations can be used to estimate the optimal set of bandgaps and operating voltages. However, it is also insightful to estimate the optimal bandgap configuration $\{Eg_{i}\}$ alone, by considering only generation rates which are independent of the applied voltage.

A first criteria was suggested in \cite{yoshida_photon_2012} that, to favor current matching in the intermediate transitions, and thus minimize radiative losses, generation rates should verify $G_{IC}=G_{IV}$. In the general case of a narrow absorption considered in this work, this criteria cannot always be met and the current matching criteria should rather be expressed by minimizing the quantity
\begin{equation}
\left|G_{IC}-G_{IV}\right|\label{eq:crit-G}
\end{equation}

Furthermore, the voltage preservation relation eq. (\ref{eq:Voc-RBSC}) can be expressed in terms of generation and recombination rates (see appendix) as:
\begin{equation}
\frac{G_{IV}G_{IC}}{R_{IV}^{0}R_{IC}^{0}}=\frac{G_{CV}}{R_{CV}^{0}}\label{eq:crit-Voc}
\end{equation}

Since these two criteria only depend on the energy gaps, the optimal values of $Eg_{IV}$ and $Eg_{IC}$ can be determined if the optimal value of $Eg_{CV}$ is known. The efficiency of the structure calculated from these criteria is compared to the efficiency estimated from brute force optimization on Figure \ref{Fig:Opt}. Several conclusions can be reached.

First, despite being approximate relations, the two criteria eq.(\ref{eq:crit-G} - \ref{eq:crit-Voc}) presented here allow to recovers extremely well the values of the brute force method. This agreement shows that most of the physics of the RBSC is contained in current matching and voltage preservation.

Second, as for voltage preservation, the validity of these criteria extends on a large range around the global optimal, allowing for the estimation of the optimal ratchet band structure $\{Eg_{IV},\,Eg_{IC}\}$ for any gap $Eg_{{\rm CV}}$ of practical use.

Third, these numerical calculations show that, for both criteria to be met simultaneously,
a non-zero ratchet shift $\Delta E$ must be considered. An important corollary of this result is therefore that a standard IBSC without electronic ratchet can not satisfy current matching and voltage preservation at the same time under un-concentrated light.

\begin{Figure}
 \centering
 \includegraphics[width=\linewidth]{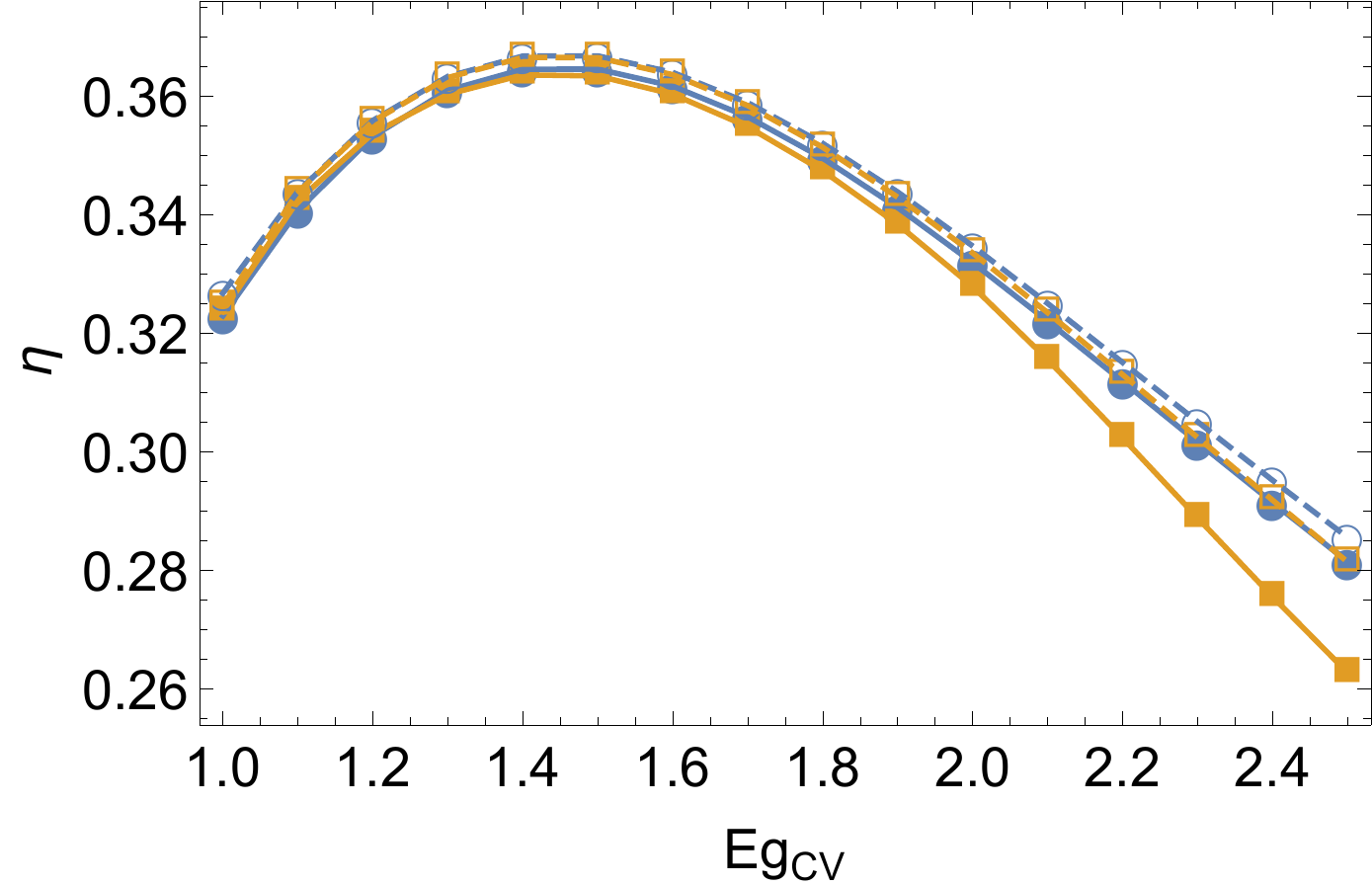}
 \captionof{figure}{Conversion efficiency as estimate by the brute force method (dashed line and empty symbols) and from criteria eq.(\ref{eq:crit-G}-\ref{eq:crit-Voc}) (plain line and filled symbols). Absorption width is taken to $\delta E = 150 \, {\rm meV}$ and $r_{{\rm rad},IC}$ to $1$ (blue circles) and $10^{-3}$ (yellow square).}
\label{Fig:Opt}
\end{Figure}

\begin{Figure}
 \centering
 \includegraphics[width=\linewidth]{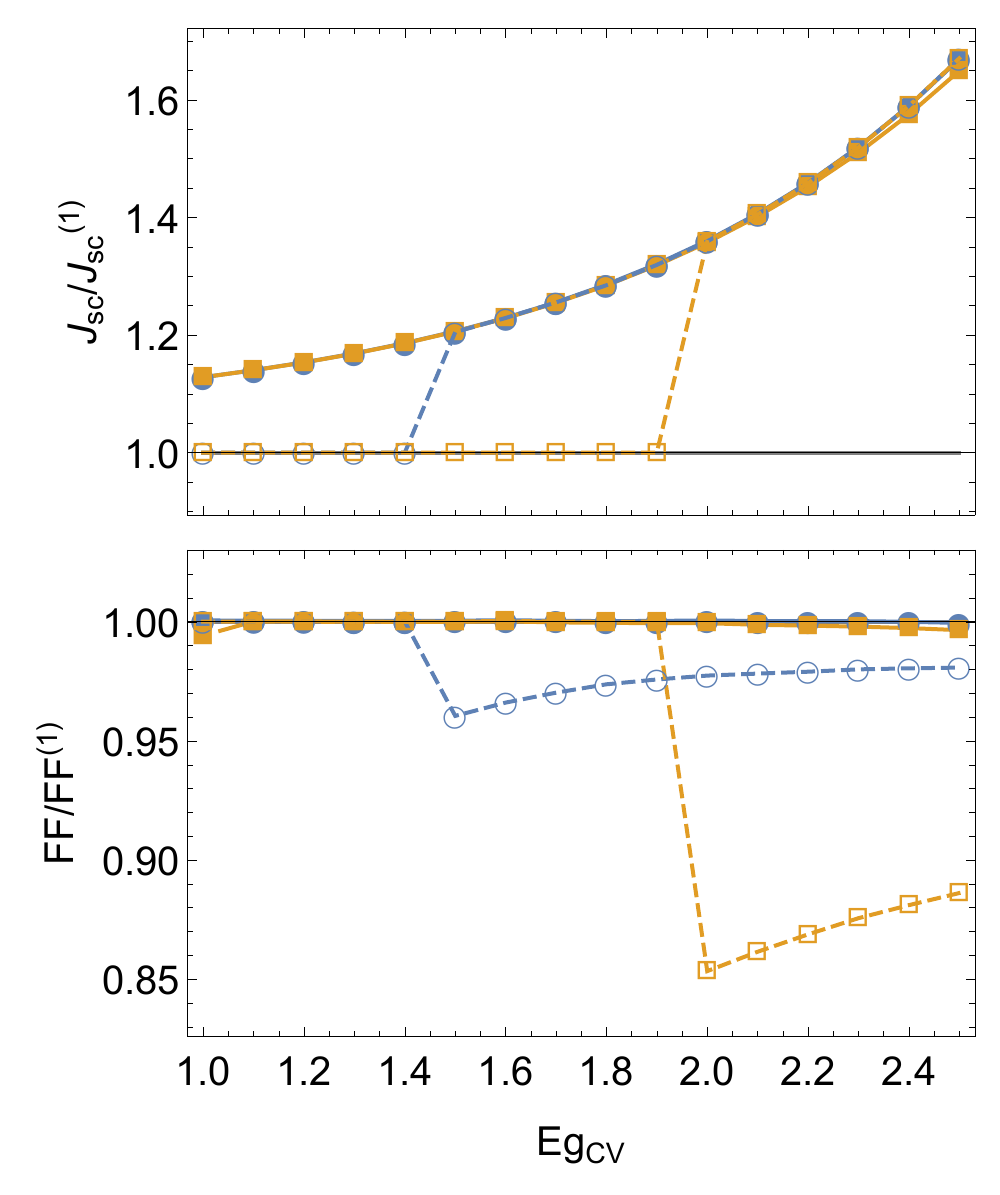}
 \captionof{figure}{Short circuit current (above) and Fill Factor (below) of RBSC (plain line and filled symbols) and IBSC (dashed line and empty symbols), normalized to that of a single junction with the same main gap $Eg_{CV}$. Absorption width is taken to $\delta E = 150 \, {\rm meV}$ and $r_{{\rm rad},IC}$ to $1$ (blue circles) and $10^{-3}$ (yellow square). For each gap $Eg_{CV}$, the band configuration of both structures are determined by optimizing the conversion efficiency. As noted in previous works \cite{delamarre_electronic_2017}, unlike for  standard IBSC, the efficiency of a RBSC is marginally affected by non-radiative losses.}
\label{Fig:Jsc-FF}
\end{Figure}
\section{The best of both worlds}

Finally, as a qualitative extension to the previous study, we estimate numerically complementary figures of merit. Figure \ref{Fig:Jsc-FF} compares compares the short circuit current $J_{\rm SC}$ and the fill factor  $FF$ reached by an optimal RBSC, IBSC and single junction sharing the same large $Eg_{CV}$. 

As in Fig. \ref{Fig:Voc}, the IBSC works as a single junction for small gaps. The intermediate transition contribution appears as a significant increase in the short circuit current, but is accompanied by a decrease of the fill factor (in addition to the voltage loss mentioned above). When non radiative losses are considered, the trade-off between the beneficial and detrimental effects of the intermediate transition prevent the system from exceeding the Shockley-Queisser limit \cite{delamarre_electronic_2017}. 

By contrast, the RBSC maintains also a fill factor very close to that of a single junction, regardless of the main gap or radiative efficiency considered, and benefits from the same current increase as an optimal IBSC. In perspective of voltage preservation demonstrated above, it thus appears that the electronic ratchet enables RBSC to benefit from the best of two worlds, and to take advantage of the best characteristics in both IBSC and single junction systems.

\section{Conclusions}
In summary, we have studied theoretically electrical properties of intermediate band solar cells featuring an electronic rachet. We have shown numerically that RBSC can reach the same open circuit voltage and fill factor as a single junction with the same gap, while benefiting from the same current increase as a standard IBSC without ratchet. To investigate the underlying physics, we have proven that the optimal RBSC is characterized by  voltage preservation and current matching between IC and IV transitions. By contrast, an IBSC without ratchet is unable to meet both criteria simultaneously and the resulting trade-off limits its conversion efficiency. This analytical result has been obtained using Lagrange multipliers, a method particularly well suited for optimization under constraints and which will certainly bring new insights on many problems in photovoltaics.

\section*{Acknowledgments}
DS and NC thank the Japan Society for the Promotion of Science (JSPS) for financial support (grant number PE16763).

\end{multicols}

\section*{Appendix}

\subsubsection*{Open-circuit voltage and Maximum power point in IBSC}

In this section, we demonstrate eq.(\ref{eq:Vm-Voc}) and (\ref{eq:crit-Voc})
used in the main text from the I-V behavior of an IBSC (with or without
ratchet). To estimate this behavior, we express the current flowing
at a bias $qV=\Delta\mu_{CV}$ as the sum of the direct and intermediate
contributions
\begin{equation}
\frac{J}{q}=\left(G_{CV}-R_{CV}^{0}e^{\Delta\tilde{\mu}_{CV}}\right)+\left(G_{IC}-R_{IC}^{0}e^{\Delta\tilde{\mu}_{IC}}\right)
\end{equation}

Owing to Kirchhoff laws eq. (\ref{eq:maille}) and (\ref{eq:noeuds}),
it is possible to estimate the recombination rate from conduction
to intermediate band
\begin{equation}
R_{IC}^{0}e^{\Delta\tilde{\mu}_{IC}}=\frac{G_{IC}-G_{IV}+\sqrt{\left(G_{IC}-G_{IV}\right)^{2}+4R_{IV}^{0}R_{IC}^{0}e^{\Delta\tilde{\mu}_{CV}}}}{2}
\end{equation}
and the current can be expressed as a function of the applied bias
only:

\begin{align}
\frac{J}{q} & =\left(G_{CV}-R_{CV}^{0}\,e^{qV/kT_{C}}\right)+\left(\frac{G_{IV}+G_{IC}}{2}-\sqrt{\left(\frac{G_{IV}+G_{IC}}{2}\right)^{2}+R_{IV}^{0}R_{IC}^{0}\left(\,e^{qV/kT_{C}}-\frac{G_{IV}G_{IC}}{R_{IV}^{0}R_{IC}^{0}}\right)}\right)
\end{align}
From this expression, it is straightforward to show that, for the
open circuit voltage $V_{{\rm OC}}$ to match that of the corresponding
single junction $V_{OC}^{(1)}=\frac{kT_{C}}{q}\log\left(\frac{G_{CV}}{R_{CV}^{0}}\right)$,
generation and recombination rates must verify
\begin{equation}
\frac{G_{IV}G_{IC}}{R_{IV}^{0}R_{IC}^{0}}=\frac{G_{CV}}{R_{CV}^{0}}
\end{equation}
hence eq.(\ref{eq:crit-Voc}). Furthermore, provided that IC and IV
recombinations are low enough, as is the case thanks to the ratchet
energy shift, we consider $\left(\frac{G_{IV}+G_{IC}}{2}\right)^{2}\gg R_{IV}^{0}R_{IC}^{0}\left(\,e^{qV/kT}-\frac{G_{IV}G_{IC}}{R_{IV}^{0}R_{IC}^{0}}\right)$
and the previous expression can be simplifed to
\begin{equation}
\frac{J}{q}\simeq\left(G_{CV}+\frac{G_{IV}G_{IC}}{G_{IV}+G_{IC}}-e^{qV/kT_{C}}\left(R_{CV}^{0}+\frac{R_{IV}^{0}R_{IC}^{0}}{G_{IV}+G_{IC}}\right)\right)
\end{equation}

It is straightforward to estimate the open circuit voltage $V_{{\rm OC}}$
 by setting $J=0$
\begin{equation}
e^{qV_{{\rm OC}}/kT_{C}}=\left(G_{CV}+\frac{G_{IV}G_{IC}}{G_{IV}+G_{IC}}\right)/\left(R_{CV}^{0}+\frac{R_{IV}^{0}R_{IC}^{0}}{G_{IV}+G_{IC}}\right)\label{eq:Voc}
\end{equation}

The maximal power point $V_{{\rm m}}$ is defined by $\left.\partial_{V}\,J\times V\right|_{V=V_{{\rm m}}}=0$,
leading to
\begin{align}
0 & =G_{CV}+\frac{G_{IV}G_{IC}}{G_{IV}+G_{IC}}-\left(R_{CV}^{0}+\frac{R_{IV}^{0}R_{IC}^{0}}{G_{IV}+G_{IC}}\right)\left(1+\frac{qV_{{\rm m}}}{k_{B}T}\right)e^{qV_{{\rm m}}/kT_{C}}
\end{align}
and the prefactor can be recognized from eq. (), resulting in the
same relation between $V_{{\rm OC}}$ and $V_{{\rm m}}$ as in a single
junction
\begin{equation}
\left(1+\frac{qV_{{\rm m}}}{kT}\right)\exp\left(\frac{qV_{{\rm m}}}{kT_{C}}\right)=\exp\left(\frac{qV_{{\rm OC}}}{kT_{C}}\right)
\end{equation}
hence eq.(\ref{eq:Vm-Voc}).

\subsubsection*{Lagrange multipliers}

In this section, we derive results presented in the main text based
on Lagrange multiplier approach. This approach consists in treating
all parameters as free parameters, and accounting for constraints
through additionnal terms included in the Lagrangian (\ref{eq:LR}).
The optimum configuration respecting these constraints is reached
when the derivative of $L_{{\rm R}}$ with respect to all parameters
and multipliers is zero. It is straightforward to show that Kirchhoff's
laws eq.(\ref{eq:maille}-\ref{eq:noeuds}) are recovered from the
derivative of the Lagrangian with respect to the Lagrange multipliers
$\lambda_{V}$ and $\lambda_{J}$ respectively.

Let us first consider the derivative of $L_{{\rm R}}$ with respect
to $Eg_{IC}$ and $Eg_{IV}$ respectively. 

\begin{align}
\partial_{Eg_{IC}}L_{{\rm R}} & =0\Rightarrow\Delta\mu_{IC}=qV_{{\rm gm},\,IC}\,{\rm or}\,\Delta\mu_{IC}=\lambda_{I}\\
\partial_{Eg_{IV}}L_{{\rm R}} & =0\Rightarrow\Delta\mu_{IV}=qV_{{\rm gm},\,IV}\,{\rm or}\,\Delta\mu_{IV}=-\lambda_{I}
\end{align}
 These relations result in four options, which we will now examine.
\begin{description}
\item [{Option~1}] $\Delta\mu_{IC}=\lambda_{I}$ and $\Delta\mu_{IV}=-\lambda_{I}$

In this case, following Kirchhoff's voltage law eq.(\ref{eq:maille}),
we find that the maximum power point is $qV_{{\rm m}}=\Delta\mu_{CV}=0$,
which is absurd.
\end{description}
To disciminate between the three remaining options, we consider derivatives
of the Langragian with respect to the quasi-Fermi levels splitting
$\Delta\mu_{IC}$ and $\Delta\mu_{IV}$, together with Kirchhoff's
current law eq.(\ref{eq:noeuds}), leading to:

\begin{equation}
\left(\Delta\mu_{IV}+\lambda_{I}\right)R_{IV}^{0}e^{\Delta\tilde{\mu}_{IV}}=\left(\Delta\mu_{IC}-\lambda_{I}\right)R_{IC}^{0}e^{\Delta\tilde{\mu}_{IC}}
\end{equation}

\begin{description}
\item [{Option~2}] $\Delta\mu_{IC}=\lambda_{I}$ and $\Delta\mu_{IV}=qV_{{\rm gm},\,IV}$

In this case, the right hand side is zero, leading to $qV_{{\rm gm},\,IV}=\Delta\mu_{IV}=-\lambda_{I}$
and we are back to option 1.
\item [{Option~3}] $\Delta\mu_{IC}=qV_{{\rm gm},\,IC}$ and $\Delta\mu_{IV}=-\lambda_{I}$

In this case, the left hand side is zero, leading to $qV_{{\rm gm},\,IC}=\Delta\mu_{IC}=\lambda_{I}$
and we are back to option 1.
\item [{Option~4}] $\Delta\mu_{IC}=qV_{{\rm gm},\,IC}$ and $\Delta\mu_{IV}=qV_{{\rm gm},\,IV}$

We are left with only this option, and will consider in the following
\begin{align}
\Delta\mu_{IC} & =qV_{{\rm gm},\,IC} \label{eq:muIC} \\
\Delta\mu_{IV} & =qV_{{\rm gm},\,IV} \label{eq:muIV}
\end{align}

\end{description}
These expressions can be used to estimate the expression of the Lagrange
multiplier:
\begin{equation}
\lambda_{I}=\frac{\Delta\mu_{IC}\,R_{IC}^{0}\,\exp\left(\frac{\Delta\mu_{IC}}{kT_{C}}\right)-\Delta\mu_{IV}\,R_{IV}^{0}\exp\left(\frac{\Delta\mu_{IV}}{kT_{C}}\right)}{R_{IV}^{0}\exp\left(\frac{\Delta\mu_{IV}}{kT_{C}}\right)+R_{IC}^{0}\exp\left(\frac{\Delta\mu_{IC}}{kT_{C}}\right)}
\end{equation}
which in turn can be used to develop $\partial_{Eg_{CV}}L_{{\rm R}}=0$
as
\begin{equation}
\exp\left(\frac{qV_{{\rm m}}}{kT_{C}}\right)\left(1+\frac{R_{IC}^{0}}{R_{IV}^{0}}\frac{\exp\left(\frac{\Delta\mu_{IC}}{kT_{C}}\right)-1}{\exp\left(\frac{\Delta\mu_{IV}}{kT_{C}}\right)}\right)=\frac{\partial_{Eg_{CV}}G_{CV}}{\partial_{Eg_{CV}}R_{CV}}
\end{equation}
hence eq.(\ref{eq:Vm-full}). 

\end{document}